\documentclass[
reprint,
amsmath,amssymb,
aps,
prl,
floatfix,
superscriptaddress
]{revtex4-1}

\usepackage{graphicx}					
\usepackage{dcolumn}					
\usepackage{bm}						
\usepackage{float}
\usepackage{gensymb}
\usepackage{multirow}

\renewcommand{\eqref}[1]{Eq.~(\ref{#1})}

\newcommand{\Figref}[1]{Fig.~\ref{#1}}

\newcommand{\threevec}[1]{\mathbf{#1}}
\newcommand{\twovec}[1]{\mathbf{#1}}

\renewcommand{\i}{\text{in}}

\newcommand{\Cu}{\text{Cu}}
\newcommand{\Co}{\text{Co}}
\newcommand{\Pt}{\text{Pt}}

\newcommand{\mhat}{\hat{\threevec{m}}}

\newcommand{\xhat}{\hat{\threevec{x}}}
\newcommand{\yhat}{\hat{\threevec{y}}}
\newcommand{\zhat}{\hat{\threevec{z}}}

\newcommand{\khat}{\hat{\threevec{k}}}
\newcommand{\kp}{\twovec{k}_{||}}
\newcommand{\kvec}{\threevec{k}}

\newcommand{\field}{\hat{\threevec{s}}}
\newcommand{\phat}{\hat{\threevec{p}}}
\newcommand{\uhat}{\hat{\threevec{u}}}

\begin{document}

\preprint{APS/123-QED}


\title{Interface-generated spin currents}


\author{V. P. Amin}
\email{vivek.amin@nist.gov}
\affiliation{
Maryland NanoCenter, University of Maryland, College Park, MD 20742
}
\affiliation{
Center for Nanoscale Science and Technology, National Institute of Standards and Technology, Gaithersburg, Maryland 20899, USA
}
\author{J. Zemen}
\affiliation{
Faculty of Electrical Engineering, Czech Technical University in Prague, Technická 2, Prague 166 27, Czech Republic
}
\author{M. D. Stiles}
\affiliation{
Center for Nanoscale Science and Technology, National Institute of Standards and Technology, Gaithersburg, Maryland 20899, USA
}


\date{\today}


\begin{abstract}


Transport calculations based on ab-initio band structures reveal large interface-generated spin currents at Co/Pt, Co/Cu, and Pt/Cu interfaces.  These spin currents are driven by in-plane electric fields but flow out-of-plane, and can have similar strengths to spin currents generated by the spin Hall effect in bulk Pt.  Each interface generates spin currents with polarization along $\zhat \times \threevec{E}$, where $\zhat$ is the interface normal and $\threevec{E}$ denotes the electric field.  The Co/Cu and Co/Pt interfaces additionally generate spin currents with polarization along $\mhat \times (\zhat \times \threevec{E})$, where $\mhat$ gives the magnetization direction of Co.  The latter spin polarization is controlled by---but not aligned with---the magnetization, providing a novel mechanism for generating spin torques in magnetic trilayers.

\end{abstract}



\maketitle



\emph{Introduction---}The Hall effect and related phenomena are pillars of condensed matter physics and are important both scientifically and technologically.  Two of these phenomena, the anomalous Hall effect \cite{AHETheoryKundt,AHETheoryPughRostoker,AHETheoryKarplusLuttinger,AHEReviewNagosa} and the spin Hall effect \cite{SHETheoryDyakonovPerel, SHETheoryHirsch, SHETheoryZhang, SHETheoryMurakami, SHETheorySinova, SHEExpKato, SHEExpWunderlich}, are particularly important in spintronics.  Both arise from spin-orbit coupling, a relativistic phenomenon that couples the spin and momentum of carriers.  They occur when an electric field drives carriers and those with opposite spins deflect in opposite directions.

In nonmagnetic heavy metals, spin-orbit scattering deflects the same amount of carriers in opposite directions, creating a transverse flow that forms a pure spin current (no charge current).  This is known as the spin Hall effect.  In contrast, ferromagnets support an imbalance of majority and minority carriers at the Fermi level, which gives rise to spin-dependent scattering rates and spin-dependent conductivities.  Thus, when majority and minority carriers are oppositely deflected, the current associated with each spin is different.  This results in both a transversely flowing spin current and a transversely-flowing charge current.  This is known as the anomalous Hall effect, because it acts like the original Hall effect but is determined by the magnetization rather than the external magnetic field.

Both effects electrically generate spin currents, but they generate spin currents with different spin polarizations.  For the spin Hall effect, the charge flow, spin flow, and spin polarization are mutually orthogonal.  Thus, an electric field along $\xhat$ generates a spin current flowing along $\zhat$ with spin polarization along $\yhat$ (as well as a spin current flowing along $\yhat$ with spin polarization along $-\zhat$).  For the anomalous Hall effect, focus has historically been on the charge current and its spin polarization has not been well studied.  While strong dephasing in ferromagnets might suggest that carrier spins tend to point along the magnetization ($\mhat$) \cite{AHEAMRTaniguchi}, competition between exchange fields and spin-orbit fields allows carrier spins to point along other directions.

\begin{figure}[b!]
	\centering
	\vspace{0pt}	
	\includegraphics[width=1\linewidth,trim={0cm 0cm 0cm 0cm},clip]{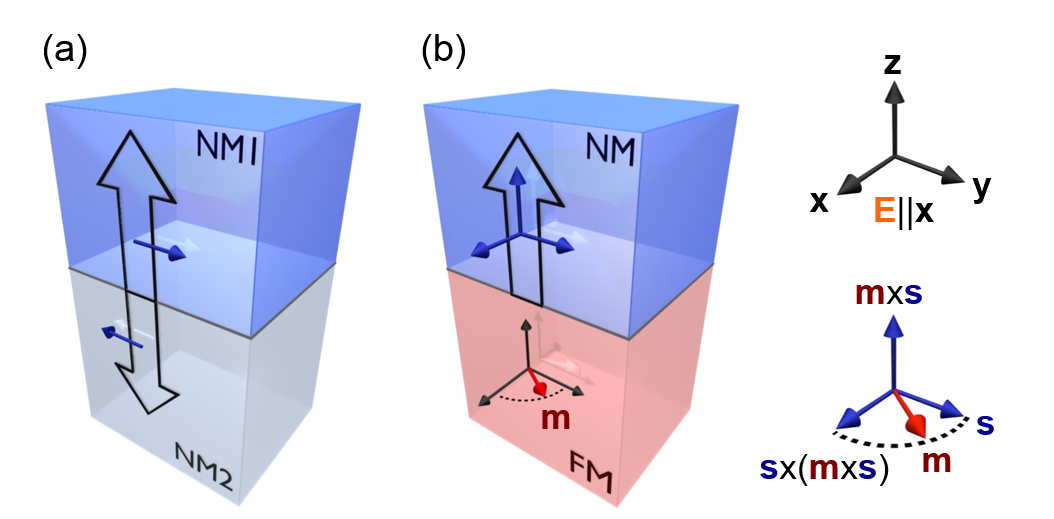}
	\caption{
	(Color online) Depiction of interface-generated spin currents at (a) nonmagnetic and (b) ferromagnet/nonmagnet interfaces driven by an electric field $\threevec{E} || \xhat$.  Block arrows show the spin flow direction ($\zhat$) and blue arrows show the allowed spin polarization directions.
	}
	\vspace{0pt}
	\label{fig:Devices}
\end{figure}

Recent work has shown that interfaces with spin-orbit coupling generate spin currents when driven by an electric field in the interface plane \cite{iSOCAminFormalism,iSOCAminPhenomenology}.  These interface-generated spin currents can flow out of the interface plane and exert spin torques on adjacent or nearby ferromagnet layers.  Like the spin Hall effect, these interface-generated spin currents convert a charge current into a transversely-flowing spin current.  Due to the reduced symmetry of the interface, the interface-generated spin current flowing out of plane can be written as
\begin{align}
\threevec{j}		&=		j_\text{f} \field		+		j_\text{p} \mhat \times \field		+		j_\text{m} \field \times (\mhat \times \field),		\label{eq:GeneralForm}
\end{align}
where the three-vector $\threevec{j}$ points along the spin polarization direction.  For high symmetry interfaces, $\field \equiv \zhat \times \threevec{E}$.  At nonmagnetic interfaces, $j_\text{p}$ and $j_\text{m}$ vanish.  At ferromagnet/nonmagnet interfaces, $j_\text{m}$ vanishes when $\mhat$ points in-plane or out-of-plane.  \Figref{fig:Devices} summarizes interface-generated spin currents.

In this paper, we compute the strength and magnetization dependence of interface-generated spin currents using first principles transport calculations.  We find that Co/Pt, Co/Cu, and Pt/Cu interfaces generate spin currents with coefficients similar to spin Hall conductivities reported in Pt and are as large as $\approx$1500 $\Omega^{-1} \text{cm}^{-1}$\footnote{We use units of conductivity for spin currents.  A spin current in units of a flux of angular momentum may be converted to units of a charge current density by multiplying by a factor of $2e/\hbar$, then to a conductivity by dividing by the electric field.}.  The spin currents injected into ferromagnetic layers dephase and create spin-orbit torques.  In this paper, we focus on the spin currents injected into non-magnetic layers, which can traverse that layer and create torques on a remote ferromagnetic layer.  In principle, these interface-generated spin currents enable field-free switching of perpendicularly-magnetized layers in ferromagnetic trilayers \cite{iSOCBaekAmin}, potentially important for the development of magnetic memories.  Evidence of torques exerted by such spin currents has been observed in NiFe/Ti/CoFeB and CoFeB/Ti/CoFeB spin valves \cite{iSOCBaekAmin} and in more complex multilayered systems \cite{SOTExpHumphries}.

\emph{Spin-orbit filtering/precession---} A simple model \cite{iSOCBaekAmin,iSOCAminPhenomenology} provides physical motivation for the  important effects found in the first principles calculations of spin currents and torques.  Based on this model, we call $j_\text{f} \field$ the spin-orbit filtering current and $j_\text{p} \mhat \times \field$ the spin-orbit precession current.  

\begin{figure}[t!]
	\centering
	\vspace{0pt}	
	\includegraphics[width=1\linewidth,trim={0cm 0cm 0cm 0cm},clip]{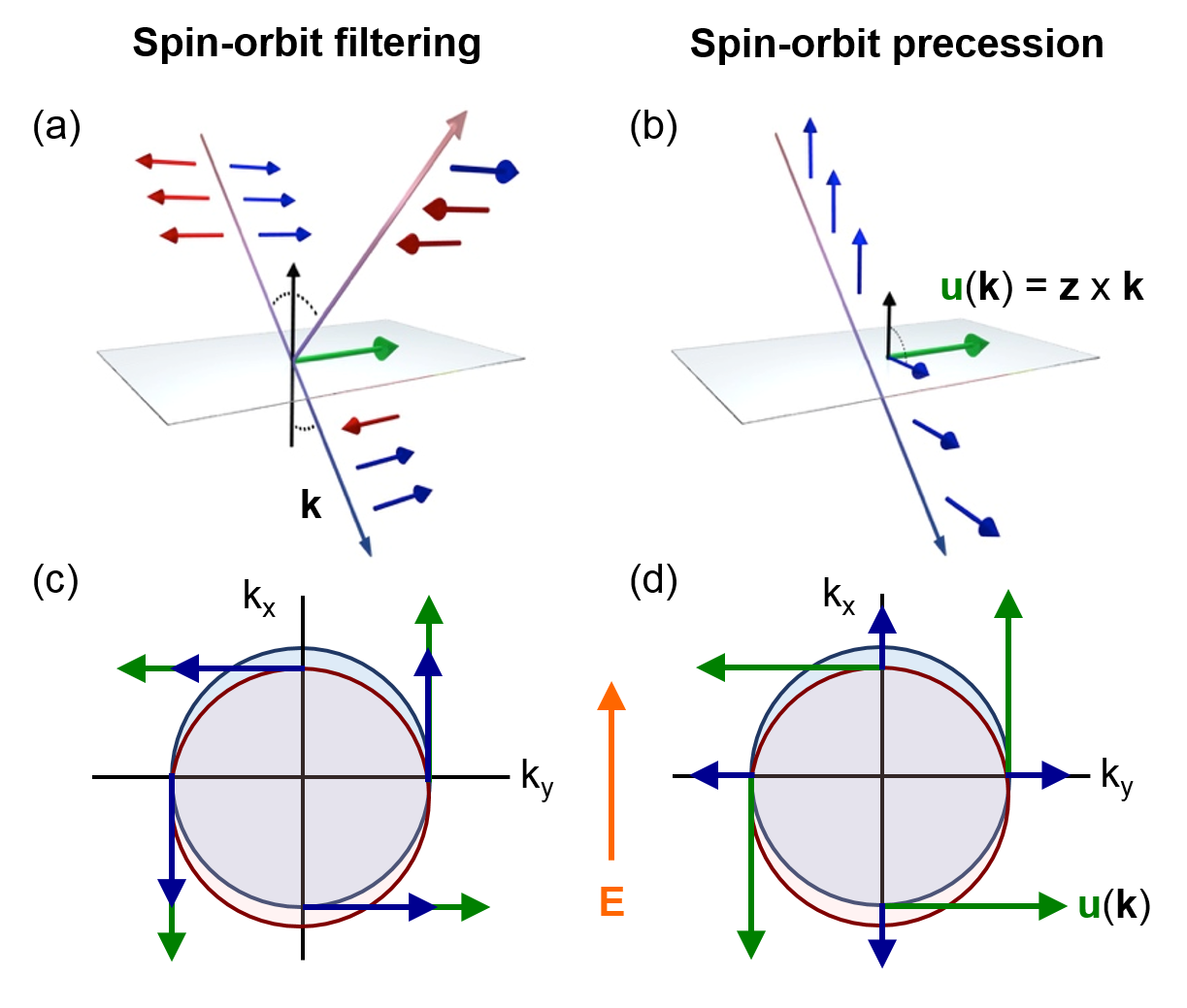}
	\caption{
	(Color online) (a) Spin-orbit filtering and (b) spin-orbit precession.  Spin-orbit filtering occurs when the interfacial spin-orbit field $\threevec{u}(\kvec)$ (green arrow) filters reflected/transmitted electrons based on spin orientation.  Spin-orbit precession occurs when electrons precess about the interfacial spin-orbit field during reflection/transmission.  (c)-(d) The transmitted in-plane spin density (blue arrows) for each incident momentum state for (c) spin-orbit filtering and (d) spin-orbit precession.  The incident states shown belong to a constant $k_z$ contour of the three-dimensional spherical Fermi surface.  The green arrows represent the momentum-dependent spin-orbit field.  Applying an electric field shifts the occupation of the incident electrons so that the transmitted electrons carry a net spin polarization.
	}
	\vspace{0pt}
	\label{fig:SOFSOPExp}
\end{figure}

Spin-orbit filtering occurs because carriers with spins parallel and antiparallel to the interfacial spin-orbit field experience different scattering amplitudes (\Figref{fig:SOFSOPExp}(a)).  Thus, reflected and transmitted carriers are spin polarized even if incoming carriers are unpolarized.  Spin-orbit filtering currents occur at nonmagnetic interfaces with spin-orbit coupling even if the bulk currents are unpolarized.  Spin-orbit precession occurs when carriers precess about the interfacial spin-orbit field while scattering off the interface (\Figref{fig:SOFSOPExp}(b)).  In the simplest models, carriers with opposite spins but the same incoming momentum precess identically while scattering, so the scattered spins remain opposite but have changed their overall orientation.  Thus, the scattered carriers are polarized only if the incident carriers are polarized, and so spin-orbit precession currents only occur if one layer is ferromagnetic.

The simple model to illustrate these processes assumes that carriers in both layers comprise free electron gases with identical, spin-independent, spherical Fermi surfaces.  If one layer is ferromagnetic, we assume the imbalance of majority and minority carriers only arises in nonequilibrium.  The interfacial potential has the form
\begin{align}
V(\threevec{r})		=		\frac{\hbar^2 k_\text{F}}{m}	\delta(z)		( u_0 + u_\text{R} \threevec{\sigma} \cdot (\khat \times \zhat) ),		\label{eq:ScatPot}
\end{align}
where $u_0$ is a spin-independent barrier, $u_\text{R}$ is the scaled Rashba parameter, $k_\text{F}$ is the Fermi momentum, and $\khat$ is a unit vector pointing along the incident momentum.  Electrons scattering from the potential in \eqref{eq:ScatPot} have the transmission amplitudes
\begin{align}
t^{\pm}(\kvec) = \frac{ i k_z/k_\text{F} }{ i k_z/k_\text{F} - ( u_0 \pm u_\text{eff}(\kvec) ) },		\label{eq:tAmp}
\end{align}
where $\pm$ label spins parallel/antiparallel to the spin-orbit field defined by $\threevec{u}(\kvec) = u_\text{eff}(\kvec) \uhat(\kvec) = u_\text{R} \khat \times \zhat$.

To compute spin currents carried by an ensemble of electrons driven by an in-plane electric field $\threevec{E}$, we determine the nonequilibrium distribution function $g_{\uparrow/\downarrow}(\kvec)$.  Here $\uparrow/\downarrow$ denote spins parallel/antiparallel to the magnetization.  For simplicity, we assume that electrons incident on the interface obey the relaxation time approximation, i.e. $g_{\uparrow/\downarrow}(\kvec) \propto E \tau_{\uparrow/\downarrow} k_x$ for $\threevec{E} = E\xhat$.  Here $\tau_{\uparrow/\downarrow}$ give the momentum relaxation times for each spin species.

While the incident electrons have no net out-of-plane flow, the reflected and transmitted electrons carry a spin current that flows out-of-plane.  In the following, we group the incident electrons into an unpolarized distribution $g^c = (g^\uparrow + g^\downarrow)/2$ and a spin-polarized distribution $g^s = (g^\uparrow - g^\downarrow)/2$.  The transmitted spin currents from the unpolarized incident electrons $\threevec{j}^t_\text{f}$ and from the spin-polarized incident electrons $\threevec{j}^t_\text{p}$ are
\begin{align}
\threevec{j}^t_\text{f}		\propto	E (\tau_\uparrow + \tau_\downarrow)	\int_\text{FS} & d\kp k_x \Big{(} |t^+(\kvec)|^2 - |t^-(\kvec)|^2 \Big{)} \uhat(\kvec),			\label{eq:SCUP}		\\
\threevec{j}^t_\text{p}	\propto	E (\tau_\uparrow - \tau_\downarrow)	\int_\text{FS} & d\kp k_x {\sf T}(\kvec) \mhat,														\label{eq:SCP}
\end{align}
where the total transmitted spin current is $\threevec{j}^t = \threevec{j}^t_\text{f} + \threevec{j}^t_\text{p}$ \cite{iSOCBaekAmin}.  Here $\kp$ denotes the in-plane momentum and ${\sf T}(\kvec)$ is a 3$\times$3 matrix that depends on $t^{\pm}(\kvec)$.  Note that $\tau_{\uparrow/\downarrow}$ correspond to the layer containing the incident electrons.  Similar expressions exist for the spin currents $\threevec{j}^r_\text{f/p}$ carried by the reflected electrons.  Further details, such as the explicit form of ${\sf T}(\kvec)$, can be found in \cite{iSOCBaekAmin}.  The total spin current is the sum of the reflected and transmitted spin currents given by $\threevec{j}_\text{f/p} = \threevec{j}^r_\text{f/p} + \threevec{j}^t_\text{f/p}$, where $\threevec{j}_\text{f} \propto \field$ and $\threevec{j}_\text{p} \propto \mhat \times \field$.  \Figref{fig:SOFSOPExp}(c)-(d) depict how the sum of transmitted spins over the incident momentum states yield this result when $\field = \yhat$.  

The spin currents $\threevec{j}_\text{f}$ and $\threevec{j}_\text{p}$ arise from the spin-orbit filtering and spin-orbit precession mechanisms introduced earlier.  The spin-orbit filtering current $\threevec{j}_\text{f}$ captures incident electrons being filtered by the spin-orbit field $\uhat(\kvec)$ during transmission.  This filtering occurs if $|t^+(\kvec)| \neq |t^-(\kvec)|$, so that spins aligned with $\uhat(\kvec)$ have higher transmission probability than antialigned spins.  The spin-orbit precession current $\threevec{j}_\text{p}$ captures incident spins oriented along $\mhat$ being rotated by the spin-orbit field during transmission.  This can be seen by showing that the matrix ${\sf T}(\kvec)$ in \eqref{eq:SCP} rotates $\mhat$ about $\uhat(\kvec)$ \cite{iSOCBaekAmin}.


In this model, the reflected and transmitted electrons only carry spin-orbit filtering and spin-orbit precession currents, so the coefficient $j_\text{m}$ in \eqref{eq:GeneralForm} vanishes.  Furthermore, the coefficients $j_\text{f}$ and $j_\text{p}$ are magnetization-independent.  However, adding an interfacial exchange interaction to \eqref{eq:ScatPot} and obtaining the resulting interface-generated spin currents numerically \cite{iSOCAminFormalism,iSOCAminPhenomenology} shows that $j_\text{f}$ and $j_\text{p}$ vary with magnetization and $j_\text{m}$ only vanishes when $\mhat$ points in-plane or out-of-plane.  This greater generality arises because the effective field $u_\text{eff}$ is magnetization-dependent in the presence of an interfacial exchange interaction.  The transport calculations presented below demonstrate that the spin currents at Co/Cu, Co/Pt, and Pt/Cu interfaces exhibit this more general magnetization dependence.


\emph{Spin torques---} Spin currents that flow into ferromagnetic layers exert dampinglike spin transfer torques of the form $\tau \propto \mhat \times (\mhat \times \phat)$, where $\phat$ equals the spin polarization direction of the spin current \cite{STTTheorySlonczewski, STTTheorySlonczewski2, STTTheoryBerger, STTTheoryRalph, STTTheoryStiles}.  Since spin-orbit filtering currents and spin Hall currents have a fixed spin polarization $\phat = \field = \zhat \times \threevec{E}$, both spin currents exert torques given by $\mhat \times (\mhat \times \field)$.  Spin-orbit precession currents have a magnetization-dependent spin polarization given by $\phat = \mhat \times \field$.  At a ferromagnet/nonmagnet interface, spin-orbit precession currents exert spin torques on the ferromagnetic layer of the form $\mhat \times (\mhat \times (\mhat \times \field)) = \mhat \times (-\field)$.  Such torques can be thought of as inciting magnetization precession about $-\field$.  In general, we may classify spin torques as \emph{dampinglike} ($\tau_\text{DL} \propto \mhat \times (\mhat \times \field)$) or \emph{fieldlike} ($\tau_\text{FL} \propto \mhat \times \field$).

In trilayers consisting of a nonmagnetic spacer sandwiched between two ferromagnetic layers, the interface-generated spin currents injected into the non-magnet can traverse that layer and exert torques on a subsequent magnetic layer, thereby coupling the magnetizations of the ferromagnetic layers.  For instance, the interface between the bottom ferromagnetic layer and the nonmagnetic spacer emits a spin current that traverses the spacer layer and exerts a spin torque on the top ferromagnetic layer.  This spin torque has the form $\mhat_t \times (\mhat_t \times (\mhat_b \times \field))$, where $\mhat_{t(b)}$ describes the magnetization of the top(bottom) ferromagnetic layer.  For the case of a fixed bottom layer with $\mhat_b = \xhat$, a free top layer with $\mhat_t = \zhat$, and an electric field $\threevec{E}~||~\xhat$, the spin-orbit precession current emitted from the bottom interface has polarization along $\zhat$.  Thus, spin-orbit precession currents can damp the magnetization towards the $\threevec{z}$-axis and therefore assist in switching perpendicularly-magnetized ferromagnetic layers \cite{iSOCBaekAmin}.

\emph{Interface-generated spin currents at Co/Cu, Co/Pt, and Pt/Cu interfaces---}First principles calculations show that interfaces significantly alter spin currents generated in bulk layers through spin memory loss \cite{SMLTheoryBelashchenko,SMLTheoryDolui} or by modifying the spin Hall angle \cite{SHETheoryWang}.  We now demonstrate the converse in realistic systems: interfaces generate spin currents that will traverse neighboring bulk layers.  To do so, we study both bulk materials (or infinite crystals) and bilayers (where each layer is a semi-infinite crystaline slab).  The materials are composed of either Co, Cu, or Pt atoms.  We use a tight-binding model fitted to first principles calculations to simulate the material systems \cite{TBZemen}\footnote{We use a multi-orbital tight binding model based on the Slater-Koster method \cite{TBSlaterKoster}, with parameters fitted to reproduce ab-initio electronic structure calculations \cite{TBShiPapaconstantopoulos} including spin-dependent on site energies. We calculate the Fermi level by requiring charge neutrality of the layers. The spin-orbit coupling from \cite{TBVisscher} is used for all atoms.  The hopping parameters at the interfaces are obtained as geometric averages of hopping parameters in neighboring atoms.} and Green's functions techniques to obtain the electronic wavefunctions, spin currents, and spin torques in the sample \cite{FPTWang,FPTWang2} \footnote{Without spin-orbit coupling, the spin torque on each principal layer is the difference between the spin currents at the boundaries of that principal layer.  In the presence of spin-orbit coupling this relation does not hold due to coupling to the crystal lattice.  To account for this, all spin torques are evaluated using the operator $J_\text{ex} \threevec{\hat{\sigma}} \times \mhat$, where the exchange field $J_\text{ex}$ is orbital-dependent and determined by the Stoner splitting taken from first principles calculations of bulk metals found in \cite{TBShiPapaconstantopoulos}.}.

To introduce an in-plane electric field (using the relaxation time approximation), we assume the nonequilibrium occupation of incoming states is proportional to $E \tau^M_\sigma v^M_{x,m\sigma}(\kvec)$.  Here, $m$ gives the spin-independent band number, $\sigma \in [\uparrow/\downarrow]$ denotes the spin state, and $M \in [\Co,\Cu,\Pt]$.  The quantity $\tau^M_\sigma$ denotes the momentum relaxation time and $v^M_{x,m\sigma}(\kvec)$ gives the $x$-velocity (determined by the electronic structure).  The momentum relaxation times are the only free parameters in this theory, and are chosen to reproduce the desired bulk conductivity and polarization in each layer.  We do not compute the perturbation to the wavefunctions caused by the electric field or include the impurity scattering potentials that drive side-jump or skew-scattering.  Thus, our results exclude the spin Hall and anomalous Hall effects in all materials.  However, spin-orbit coupling modifies the bulk Co eigenmodes in a magnetization-dependent way, thus capturing the anisotropic magnetoresistance/planar Hall effect \cite{AMRExpThomson, AMRTheoryMcGuirePotter, TAMRExpGould, ChantisTAMR, AHEReviewNagosa, BauerSpinCaloritronics, TAMSTheoryAmin, TAMSTheoryCarlos}.  


\begin{figure}[t!]
	\centering
	\vspace{0pt}	
	\includegraphics[width=1\linewidth,trim={0cm 0cm 0cm 0cm},clip]{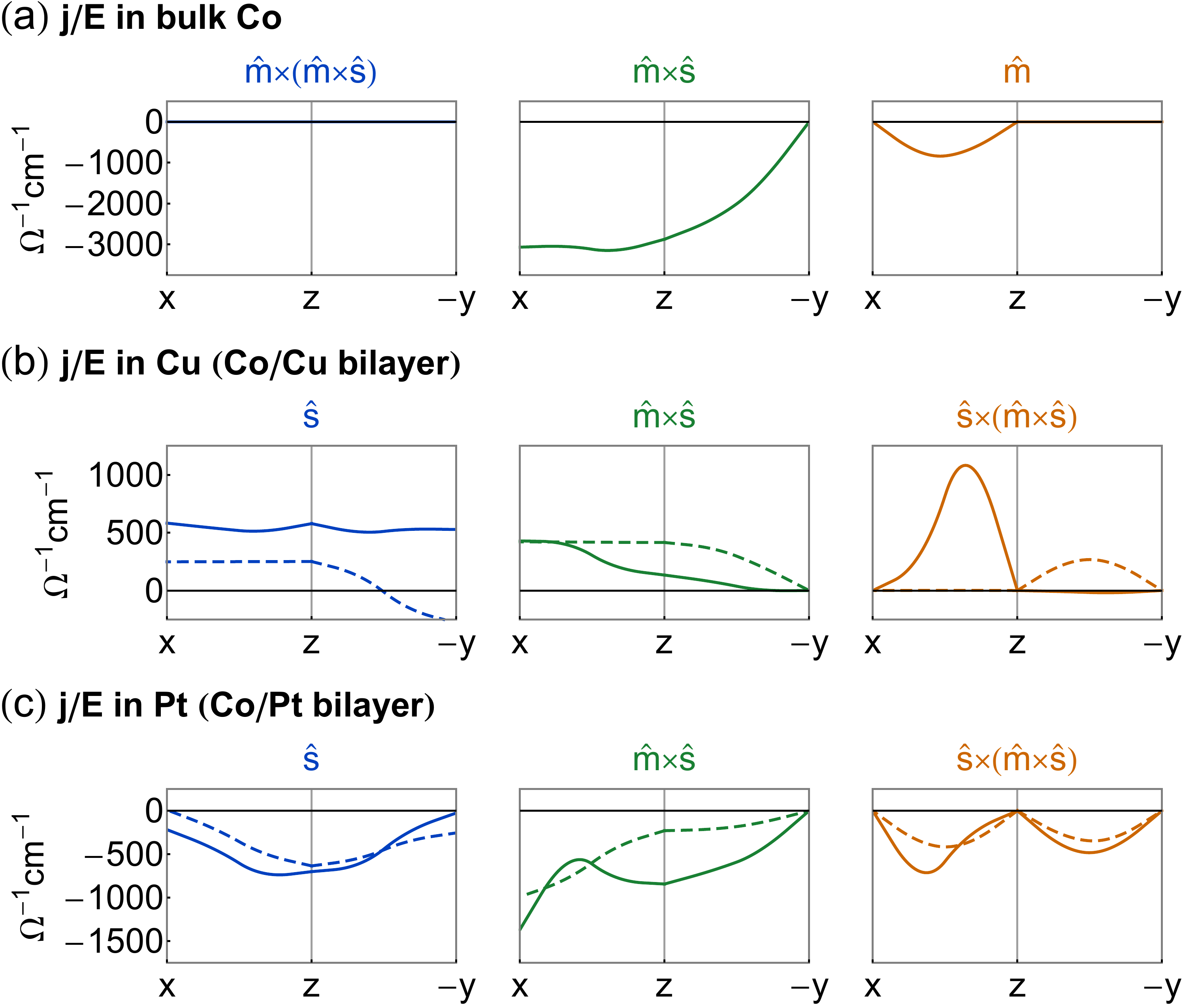}
	\caption{
	(Color online) Magnetization dependence of spin currents ($\threevec{j}$) flowing out-of-plane ($\zhat$) generated by an in-plane electric field ($\threevec{E}||\xhat$).  The magnetization is swept along the $x/z$ plane followed by the $z/y$ plane.  (a) Bulk spin currents in Co, whose origin is closely related to the planar Hall effect.  (b)-(c) Spin currents in Cu and Pt layers (averaged over 14 monolayers) within the Co/Cu and Co/Pt bilayers.  The solid curves give the total spin currents while the dashed curves give purely interface-generated spin currents (obtained by removing spin-orbit coupling from the Co layers).
	}
	\vspace{0pt}
	\label{fig:MagDep}
\end{figure}

Bulk currents arising from the planar Hall effect in Co complicate the analysis of interface-generated spin currents.  Therefore we first discuss charge and spin currents in bulk materials with flow transverse to the electric field.  As expected in the absence of the spin Hall and anomalous Hall effects, bulk Pt and Cu do not generate any transversely-flowing charge or spin currents.  However, bulk Co generates a tranversely-flowing charge current with a magnetization dependence consistent with the planar Hall effect.  This charge current is accompanied by transversely-flowing spin currents polarized along $\mhat$ and $\mhat \times \field$.  \Figref{fig:MagDep}(a) shows the magnetization dependence of these bulk spin currents.  The former spin current is expected in ferromagnets while the latter is allowed by symmetry but not typically considered.  Although the spin polarization of the latter spin current is misaligned with the magnetization, we find it exerts no spin torques and is continuous across each monolayer.  All bulk-generated spin currents (flowing transverse to the electric field) vanish in the absence of spin-orbit coupling.  


\begin{figure}[t!]
	\centering
	\vspace{0pt}	
	\includegraphics[width=1\linewidth,trim={0cm 0cm 0cm 0cm},clip]{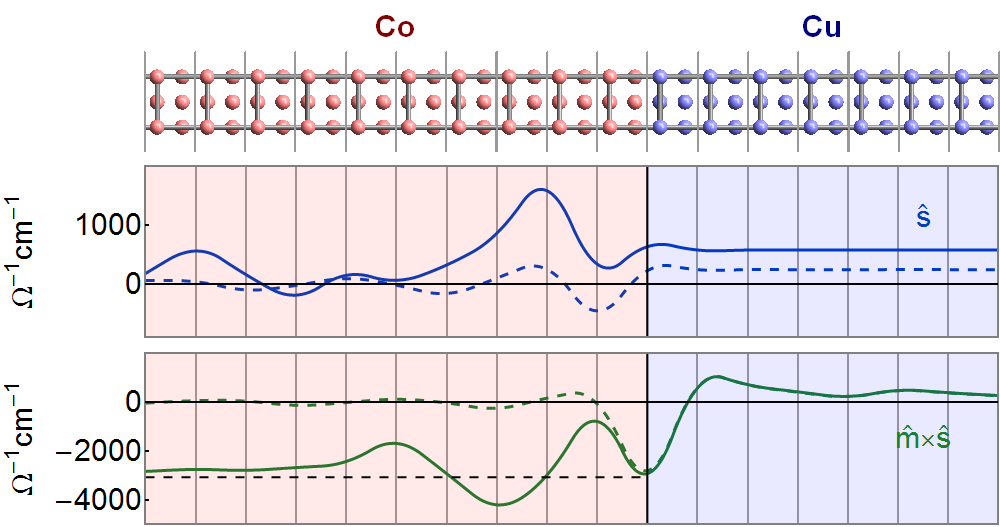}
	\caption{
	(Color online) Plot of the spin currents flowing out-of-plane ($\zhat$) driven by an in-plane electric field ($\threevec{E}||\xhat$) between each principal layer (two monolayers) in the Co/Cu bilayer (solid curves).  Artificially removing spin-orbit coupling in Co eliminates the bulk contribution from that layer, giving purely interface-generated spin currents (dashed curves).  The black horizontal dashed line gives the expected bulk contribution (taken from the bulk Co simulation).  
	}
	\vspace{0pt}
	\label{fig:LayerPlots}
\end{figure}

Next, we discuss the spin currents in bilayers that are driven by an in-plane electric field but flow out-of-plane.  In accordance with the toy model described above, the Pt/Cu bilayer (not shown) only generates spin currents polarized along $\field$ while the Co/Cu and Co/Pt bilayers generate spin currents polarized along $\field$, $\mhat \times \field$, and $\field \times (\mhat \times \field)$.  However, bulk spin current generation in the Co layers also contribute to these spin currents (discussed above).  To isolate the interface-generated spin currents, we artificially remove spin-orbit coupling in the Co layers.  The results are shown by the dashed curves in \Figref{fig:MagDep}(b)-(c) and \Figref{fig:LayerPlots}.  Even with spin-orbit coupling eliminated in the Co layers, the interface-generated spin currents that escape into Cu and Pt are not signficantly reduced.


In most devices utilizing spin-orbit torques, the spin Hall effect is thought to be intrinsic and not vary with the electron lifetimes.  Thus, the ratio of the spin current flowing out-of-plane to the charge current flowing in-plane is largest for low conductivity materials.  Interface-generated spin currents scale with electron lifetimes (which are monotonically related to the conductivity) so the same ratio is largely independent of conductivity.  These spin currents are therefore more likely to be important in high conductivity samples.  The interface-generated spin currents in Co/Cu are comparable to measured Pt spin Hall conductivities ranging from $\approx$1300 $\Omega^{-1} \text{cm}^{-1}$-1900 $\Omega^{-1} \text{cm}^{-1}$ \cite{SHEExpSagasta} and to theoretical estimates of 1300 $\Omega^{-1} \text{cm}^{-1}$ \cite{SHETheoryTanaka} and 1600 $\Omega^{-1} \text{cm}^{-1}$ \cite{SHETheoryWang}.  The values generated by Co/Pt and Pt/Cu interfaces are even larger and fall within the range of the Pt spin Hall conductivities reproduced above, although larger estimates of $\approx$3000 $\Omega^{-1} \text{cm}^{-1}$ have been obtained \cite{SPExpSanchez,SHEExpNguyen}.  For Co/Pt bilayers, this suggests that strong competition might exist between the spin Hall effect and interface-generated spin currents.

%




We have demonstrated that Co/Cu, Co/Pt, and Pt/Cu interfaces driven by an in-plane electric field generate spin currents that flow out-of-plane ($\zhat$).  All three interfaces generate spin currents polarized along $\field = \zhat \times \threevec{E}$.  For the magnetic bilayers, both the interface and the bulk ferromagnetic layer additionally generate spin currents polarized along $\mhat \times \field$ and $\field \times (\mhat \times \field)$.  These additional spin currents could enable field-free switching in magnetic trilayers where the free ferromagnetic layer has perpendicular magnetic anisotropy.  This family of bulk- and interface-generated spin currents present a novel mechanism to generate spin torques in magnetic heterostructures.

\begin{acknowledgments}
The authors thank Kyung-Jin Lee, Byong-Guk Park, Xin Fan, and Paul Haney for useful conversations and Robert McMichael and Justin Shaw for critical readings of the manuscript.  VA acknowledges support under the Cooperative Research Agreement between the University of Maryland and the National Institute of Standards and Technology Center for Nanoscale Science and Technology, Award 70NANB14H209, through the University of Maryland.
\end{acknowledgments}

\bibliography{apssamp}


\end{document}